# ConnectiCity, augmented perception of the city

Salvatore Iaconesi

La Sapienza University of Rome

ISIA Design Florence

Rome University of Fine Arts

IED Rome

salvatore.iaconesi@artisopensource.net

Oriana Persico

La Sapienza University of Rome

oriana.persico@gmail.com

**Abstract**

As we move through cities in our daily lives, we are in a constant state of transformation of the spaces around us.

Information, encounters, sounds, to-do lists, things which enter our field of vision recalling memories and thoughts, people to contact, images: all of these things are just a small part of the set of stimuli which push us to re-interpret urban spaces in novel ways.

On top of that, the ways in which we constantly personalize the spaces which we traverse and in which we perform our daily routines is also able to communicate a wide array of truly significant information about our emotional states, our working methodologies, about our knowledge, skills, cultural backgrounds, desires, visions and perception of our cities, shops, parks, offices, houses, schools, as suggested in among many others.

The form and essence of urban space directly affects people's behavior, describing in their perception what is possible or impossible, allowed or prohibited, suggested or advised against.

We are now able to fill and stratify space/time with digital information layers, completely wrapping cities in a membrane of information and of opportunities for interaction and communication.

Mobile devices, smartphones, wearables, digital tags, near field communication devices, location based services and mixed/augmented reality have gone much further in this direction, turning the world into an essentially read/write, ubiquitous publishing surface.

The usage of mobile devices and ubiquitous technologies alters the understanding of place.

In this process, the definition of (urban) landscape powerfully shifts from a definition which is purely administrative (e.g.: the borders of the flower bed in the middle of a roundabout) to one that is multiplied according to all individuals which experience that location; as a lossless sum of their perceptions; as a stratification of interpretations and activities which forms our cognition of space and time.

In our research we investigated the possibilities to use the scenario which sees urban spaces progressively filling with multiple layers of real-time, ubiquitous, digital information to conceptualize, design and implement a series of usage scenarios.

The possibility to listen to the ideas, visions, emotions and proposals which are expressed each day by citizens – either explicitly or implicitly by the ways in which they use their cities, workplaces, malls... – suggests the emergence of positive scenarios.

It is possible to create multiple layers of narratives which traverse the city and which allow us to read them in different ways, according to the different strategies and methodologies enabling us to highlight how cities (through their citizens or even on their own, expressing through sensors) express points of view on the environment, culture, economy, transports, energy and politics.

These methodologies for real-time observation of cities can be described as a form of "ubiquitous anthropology", based on the idea that we can take part in a networked structure shaped as a diffused expert system, capturing disseminated intelligence to coagulate it into a framework for the real-time processing of urban information.

**Introduction**

We live in a constant state of re-programming of the spaces around us (Porteous, 1976).

As pointed out by Krupat and Guild (1980), Nasar (1989) and Scheiberg (1990) the ways in which we reinterpret and personalize spaces effectively convey important information about our emotional states, working methodologies, knowledge, skills, cultural backgrounds, desires and vision. It is a pragmatic manifestation of the ways in which we perceive our living environments, a constructivist act of world-making.

These processes have direct impact on the transformation of spaces, which are visible at the levels of households, communities, neighborhoods and entire cities.

"In the course of time every section and quarter of the city takes on something of the character and qualities of its inhabitants. Each separate part of the city is inevitably stained with the peculiar sentiments of its population." (Gottdiener, 1994)

If we observe this scenario from a different point of view, it is also clear how the form and essence of urban space directly affects people's behavior, describing in their perception what is possible or impossible, allowed or prohibited, suggested or advised against (Horton & Reynolds, 1971).

The wide availability and accessibility of ubiquitous technologies substantially radicalizes all these tendencies.

Our experience of the contemporary world is characterized by the presence of a ubiquitous digital membrane (Zook & Graham, 2007), represented and accessed by all those technologies and networks whose wide availability and accessibility allows us to fill space/time with digital information and with opportunities for interaction, interrelation and communication (Zeisel, 2006; McCullogh, 2004; Fattahi, 2009).

We have turned this observation into a research in which multiple disciplines across arts, design, architecture, anthropology and other sciences and disciplines interrelated trying to grasp and understand human beings' continuous state of re-interpretation of the world, to infer information, suggestions and visions about the ways in which people transform their reality, in both plan and action.

A set of objectives has been set forth in the process:

- to gain a better understanding of human presence in contemporary urban spaces;
- to understand the ways in which it is possible to visualize the state of re-programming of the city enacted by human beings, transforming this continuous effort in re-interpretation of space into an accessible tool;
- to observe in real-time the digital discussions which take place in cities, to both transform them into a component of the ubiquitous information landscape of urban spaces and to understand the emotional approaches, themes and issues which emerge from human perception of the city, constantly generating insights on issues related to ecology, mobility, land use, need for services and infrastructures, sense of place, definition of emergent boundaries, attention groups, health, safety;

- to raise awareness about the ways in which different cultures, languages, backgrounds, religions, nationalities and political orientations interpret the places of the city;
- create methodologies, for all actors involved, to transform these possibilities into tools for awareness and consciousness about the expression of needs and emotions of people, for ethical, sustainable, participatory policies, plans, businesses, initiatives, processes;
- promote choral initiatives, engaging citizens, organizations and institutions.

**Transforming the sense of place**

Portable devices transform our experience of space/time.

As described by Du Gay (1997) Sony Walkman's design powerfully introduced the possibility of being in two places at once through the personalized sounds playing through our headphones, creating a powerful conjunction between physical space and the imaginary space created by the music.

Devices like the Sony Walkman allow us to traverse urban spaces – with their cognitive, aesthetic and moral significance – and to benefit from the use of a critical tool in the management of our space and time, in the construction of boundaries around ourselves, and in the creation of sites of fantasy and memory (Bull, 2000).

Mobile devices, smartphones, wearables, digital tags, location based services and mixed/augmented reality have gone much further in this direction, turning the world into an essentially read/write, ubiquitous publishing surface (Iaconesi, Persico, 2011).

The usage of mobile devices and ubiquitous technologies alters the understanding of place (Wilken, 2005).

As Morley describes: "The mobile phone is often understood (and promoted) as a device for connecting us to those who are far away, thus overcoming distance – and perhaps geography itself" (Morley, 2003).

This is one of the ways in which mobile technologies alter our perception of time and space, compressing them and, thus, redefining our possibilities to interconnect and relate to objects, processes, places and people. A few paragraphs after this first remark Morley continues by stating how the mobile phone fills "the space of the public sphere with the chatter of the earth, allowing us to take our homes with us, just as a tortoise stays in its shell wherever it travels".

This modality represent a direct, personalized intervention into the design of space, in both its form and function, creating a definite shift in the definition of (urban) landscape: from a purely administrative one to one which is multiplied according to all individuals which experience that location; a lossless sum of their perceptions; a stratification of interpretations and activities which forms our cognition of space and time.

When the two previous observations are observed from the points of view expressed in the analyses of Clément (1999), Eberhard (2009) and Farina (2010), it is possible to imagine an interesting methodological approach: the perception of space/time as a continuous, emergent, fluid, recombinant stratification of analog/digital information/communication/interaction.

The possibility to access these multiplied definitions of space alter our own perception of it, allowing us to integrate into our own cognitions coming from cultures, backgrounds and symbolical apparatuses which are potentially completely different from ours, in ways that are easily accessible from devices which we hold in our pockets.

De Kerckhove in 2001 suggested the augmentation of architecture, to include the concepts created for the World Wide Web, thus expanding our possibilities for awareness and consciousness through the wide and ubiquitous availability of multiple sources of information which are hyperlinked to the physical elements of our reality.

Operating in this direction, it is possible to imagine and design a form of disseminated intelligence which can be coagulated in multiple ways by actors traversing cities and using mobile devices to enact novel forms of reading of spaces, symbols and configurations, moving fluidly across digital and physical domains.

In 2002 Green investigated this approach from the point of view of time: the emergence of new temporalities through the usage of mobile devices, at institutional, social and subjective levels, through the evaluation and interpretation of extensive ethnographical data sets. What emerged is that mobile devices act as spatial/temporal mediators, exposing alternative perceptions and behaviors in human beings using them and, thus, proposing different usage grammars for spaces and timeframes.

This concept was translated from time to space in Berry and Hamilton's research (2010) by analyzing mobile devices usage on trains: "public places and spaces are being transformed into hybrid geographies through the introduction of new spatial infrastructure". The use of these devices enables people to systematically re-program their surroundings according to the opportunities offered by their possibility to access ubiquitous technologies and networks.

**ConnectiCity, methodology**

ConnectiCity is an arts/science meta-project which investigates the possibilities offered by the progressive availability of real-time, ubiquitous, digital layers of information, to design and implement a series of prototypes which would pursue the following goals:

- to create a set of tools allowing
    - to capture in real-time various forms of city-relevant user generated content from a variety of types of sources, including social networks, websites, mobile applications,
    - to interrelate information to the territory (Zook & Graham, 2007; Goodchild, 2010) using Geo-referencing, Geo-Parsing and Geo-Coding techniques (Lieberman, 2011; Quin, 2010; Leidner 2011; Shi & Barker, 2011), and
    - to analyze and classify information using Natural Language Processing and Named Entity Recognition techniques to identify users' emotional approaches, forms of expression, topics of interest, discussion graphs, networks of attention and of influence, trending issues, evaluations of satisfaction, well-being and happiness, and other forms of expression (using techniques designed by taking into account the many researches of this kind which have been performed over these last few years, including fundamental contributions which have been adopted from Boulos & Sanfilippo (2010), Abe & Inui (2011), Gentile & Lanfranchi (2011) and some important ones also coming from the other sources listed in the references);
- to imagine initiatives through which this information allows central administration and individuals to come together, under the form of a peer to peer ecosystem in which each subject is an informed, aware agent, thus describing novel forms of governance and decision-making processes (Snyder, 2003; Snyder, 2006; Davis et al, 2006)
- to reflect onto the life and expressions of cities and of their inhabitants, to identify new policies, new sustainable, ethical business models, urban planning processes, grass-roots initiatives, operative models;

- to use the insights provided by models such as the living labs and other user-centric innovation processes (including contributions from Alexander, Ishikawa et al, 1999; Alexander, 1999; Salingaros, 1999; Salingaros 2000; Schaffers, 2011; Mulvenna et al, 2010; Pallot et al, 2010) in the creation of novel practices for citizens, organizations and administrations;

- to reflect on the themes of cognitive accessibility for this kind of information, analyzing visual and multi-modal representation and interaction metaphors that would allow to maximize the effectiveness and ease of use and understanding of the complex information scenarios produced by the possibility to capture and ubiquitously display large quantities of geo-located, realtime information layers coming from multiple sources (Tufte, 1997; Cleveland & McGills, 1984; Spence, 2001; Wünsche, 2004);

- to confront with validation models that would allow to assess the quality, relevancy and reliability of harvested data, affected by information noise, digital-divide related issues (e.g.: not all citizens use social networks or imagine that they can use them to express opinions about their city), interpretation errors;

- to imagine and implement an openly accessible service layer based on widely accepted open standards that would allow other agencies to use the technologies and processes to easily design and build other systems and services, across disciplines and sciences;

- to reflect on the new models for identity, privacy, ethics and on the new possible emerging definitions of public and private space.

## Results

The ConnectiCity project is an on-going process started in 2008. Since then a continuous refinement of the methodologies and technologies has allowed for the creation of several prototypes which implement the concepts conceived in the investigation phase.

*Rel:attiva presenza*

The first prototype was designed in Mexico City at the Franz Mayer Museum, and it was titled "rel:attiva presenza". The occasion for this project was the presentation of the paper "architettura rel:attiva" at the Seventh International Meeting on the Revitalization of the Historical Centers, focused on the idea of architecture as mediator of the historical and contemporary city.

Rel:attiva presenza vas designed as a video projection mapping and of a sound environment in the cloister of the Italian Cultral Institute in Mexico City, in the Coyoacan neighborhood. The video projection was created by assembling video footage and images from different epochs, describing the mutation of the neighborhood across the years, starting from the beginning of the century. The images and footage were assembled together with geographical representations of the evolution of the territory and of the land use in the neighborhood. The resulting visual narrative constituted a sort of conceptual time-lapse video, in which the life of the neighborhood was shown in its evolution. The sound environment was assembled by manually harvesting field-recordings in the neighborhoods streets and markets, collecting dialogues, typical noises, sounds of transit, mobility, transport, commerce, chit-chat, voices in bars and restaurants.

The installation was proposed as a novel way to stratify the neighborhood's history into an accessible, narrative form. By looking at and listening to rel:attiva, the history of the place could be experienced along multiple points of view, in its evolution towards its present condition. Images and sounds were completely "user-generated", as they had been produced by long-time inhabitants of the place, just as the voices and sounds collected using the location's daily life.

The process designed and produced for Rel:Attiva Presenza can be thought of as a practice of archival of the perceptions, experiences and narratives of the people who live in the territory, and as a research into their accessibility. The prototype was exhibited under the form of an architectural intervention in the Coyoacan neighborhood, transforming surfaces into screens which acted as an accessibility layer for the history and emotions of the inhabitants.

The results of this first experience deeply inspired the following ones.

*The Atlas of Rome*

The following prototype created for the ConnectiCity project was, in more than one way, a direct extension of the first one.

A 35 meter long architectural projection and sound environment was created for Rome's "Festa dell'Architettura" (Architecture Fair), organized by the City Administration together with the Italian Order of the Architects in the enormous entrance corridor of the ex-Mattatoio (ex Slaughterhouse) in the Testaccio neighborhood of the city.

The Atlas of Rome's purpose was to portray in real-time the evolution of the visions, desires and actions created by architects, institutions, operators and citizens onto the city of Rome on a series of fundamental themes such as culture, creativity, education, urban planning, commerce, arts, security, health and, in general, 16 information domains which had been classified to describe the overall wellness of the city.

A complex activity was set-up in the organization of the project:

- a real-time information harvesting scheme was created to capture information in real-time from the following sources:
    - institutional and professional information sources were identified, such as blogs, websites, news feeds about the city of Rome and relevant to the chosen themes
    - relevant user accounts were identified on social networks, among those citizens, professional operators, members of the institutions, museums, art galleries, spaces for creativity and entertainment, social aggregation points, active communities, to continuously benefit from relevant updates on the chosen themes;
- harvesting took place using a selection of techniques, involving both automatic processes (RSS feed parsing, micro-formats, public API usage, authorized web-scraping, database connections, import of structured data in a variety of formats) and manual ones (such as in the case of those organizations which sent us press releases to be added into the system);
- information was parsed using Natural Language Analysis (Hanks & Pustejovsky, 2005; Tuulos & Tirri, 2004), to classify information according to the selected topics;
- information was then geo-referenced either by using the coordinates provided by the information source (for example when the information source directly corresponds to a specific place, such as in the case of museums) or extracted, whenever possible, by Geo-Parsing schemes, which was performed by using a large database of Named Entities with a geographical connotation, including the names of streets, malls, cinemas, museums, landmarks, neighborhoods, common alternative names of places, pubs, bars, shops, stores, gyms, and other dozens of types of locations for which names could be identified in the text of the harvested content ( a multi-modal text-matching engine compared the strings in multiple ways for similarity and for the textual context in which the identified words were found, to be able to filter out most false-positive results, and obtaining a correctness of about 97%);
- a series of direct input channels were created to accept content (text, images and videos) from citizens using mobile devices and a series of multitouch kiosks which were set-up in various areas of the city.

Collected information was shown on the 35 meter wide surface using a Processing application which was created to synchronize the visuals onto the 8 projections which were needed to cover the whole surface. A series of different information visualizations were designed to convey information according to different metaphors. Some where dedicated to aggregating information according to themes, time-frames and types of activities which they described.

The one which resulted in substantially different results was a geographical visualization in which color-coded circles represented the various elements of information according to their relative position (it is possible to imagine the layout of the points of interest by imagining to place them onto the map and then by removing the map from underneath them, leaving them in the correct relative positions). Points were connected by similarity: two points on the visualization were connected if they were relevant to the same themes.

In this visualization a novel form of geography emerged, constructed through the activities which take place in real-time in the city of Rome. A geography which is not made from buildings and roads, but through emotions, desires, actions and visions of the agencies which operate on the territory. This kind of emergent geography has proven to be effective in implementing peculiar readings of the city, in which the behavior of people, individually or through their organizations, describes the city, describing forms, aggregations, coherences and incoherences.

The analysis of this representation has been of fundamental value in gathering the insights which were used to create the following prototypes of the ConnectiCity project.

One peculiarity of the implementation of the Atlas of Rome was the possibility for passer-by's to interact in real-time with the installation.

Using their mobile phones (through an iPhone App) or a series of multitouch screens, visitors could interact with the part of the projection that they had in front of them. The position of the multitouch terminal and the geographical coordinates identified by the iPhone application allowed the system to understand which part of the projection to activate, eventually alerting the user that the area was currently being used by another visitor, and suggesting to moving slightly to the left or right to obtain a free projection area.

Using the touch interfaces (on the phone and surface) people could navigate detailed versions of the content. By touching interface elements including the lists of categories, users could choose which bit of information they wised to experience, and a large popup contextually appeared in front of them onto the projection, showing texts, videos and images with an easy, touch enabled interface which could be used to navigate through the various elements.

This immediate responsiveness of such a large scale projection proved to produce radically positive effects on visitors: the fact that a large-scale architectural surface was actually responding in real-time to their interactions powerfully combined with the tangible effect of having own information published onto the projection, by contributing points of view, information and visions about the chosen themes. The combined effect of being able to both contribute and interact had a distinct empowering effect on people, who spontaneously started to discuss possible uses for this kind of system in areas such as participatory urban planning, policy making and decision-making at city level.

*ConnectiCity Neighborhood Edition*

The system created for the Atlas of Rome was also implemented in a smaller scale, dedicated to provide novel scenarios for the life of neighborhoods.

An Urban Screen was designed to capture in real-time the social network conversations which could be identified as originating from within the territory of the neighborhood. For this purpose, the

Twitter, Flickr and FourSquare social networks were used, thanks to the accessibility of their geographical features, both on the side of the user publishing information, both for the systems which are supposed to read such information.

Harvested information was processed using the same, yet evolved, strategies described for the Atlas, and were shown on the Urban Screen using a simple, minimal interface in which large, black dots represented the single contributions, appearing onto the screen and connected to the edges of the screen, in which a textual representation of the content was presented. On top of that two or more dots were connected on the visualization when they represented messages dealing with the same topic or if they represented direct interactions (e.g.: re-tweets and comments).

The immediateness of the interface, allowing passer-by's to read the content and to immediately understand its context by analyzing connections, proved to be truly effective in stimulating novel forms of social, territorial interaction. People actually stopped to read the ongoing conversations, trying to identify the people behind the social network nicknames. Many times identification happened, producing enthusiastic results and creating in people the immediate awareness about the possibility to contribute to the information landscape of their neighborhood, and some people actually pulled out their smartphones and immediately started answering tweets and comments, to verify if they would actually show up in the interface.

Discussion did benefit from different levels of attention, ranging from topics related to sports and entertainment, but also engaging current news items and focal issues for the neighborhood's territory.

Most people had no problem in identifying the possibility to use such systems in terms of activating participatory processes which could create value for their neighborhood. Scenarios for self organization and coordination were imagined by most people, which imagined using the urban screen as a sort of public billboard in which to perform numerous types of coordinated actions among the residents. Some people also identified more complex usage scenarios, in which multiple types of urban screen could be imagined for different purposes, such as citizen awareness, coordination and activation, general chit-chat, practical information, requests for help and also various types of "time-bank", in which neighborhood inhabitants could exchange services among themselves.

*VersuS, Rome October 15$^{th}$*

The possibility to harvest information in real-time from cities using user generated content on social networks was used in occasion of the first instantiation of the VersuS project, part of the ConnectiCity research.

The first prototype was created in occasion of the protest which took place in the city of Rome on October 15$^{th}$ 2011.

The protest took place in the city under the form of a march authorized by the City Administration, as one of the events which were created internationally in occasion of the October 15$^{th}$ event organized worldwide by the "Occupy" movement.

In the city of Rome, the peaceful protest quickly degenerated into violence, with multiple groups of activists engaging fights with police forces which devastated large parts of the city centre, causing injuries and damage.

The harvesting component of the VersuS system was created to collect as many social network conversations as possible which were taking place during the protest in the city of Rome. Focus was placed on Facebook, Twitter and Flickr social networks, and a limited set of resources were also dedicated to FourSquare and Google+.

Different social networks were observed using different techniques. For example, Twitter streams were easily captured by using the publicly available API, as was the case of Flickr, in which public APIs allow to capture activity taking place in a specified geographical bounding box. A different technique was used to engage activity generated on Facebook: a preliminary analysis performed using the search facilities provided by the Open Graph protocol and Facebook's implementation (titled Graph API) allowed researchers to identify more than 60000 user profiles among the ones whose public "home location" (the place which users specify as being the one they live in) was described as being "Rome" or one of the hundreds of smaller cities within 80km distance of the city centre, which were merged to the about 80000 profiles which explicitly mentioned the protest in Rome during the two days before October 15$^{th}$. The list of "friends" of these users was collected as well. Duplicates were removed from the overall list, arriving to a total of more than 160000 users which were considered to be relevant to the required observation.

All identified sources of information were provided with a procedure to capture their online activities for the whole duration of the protest. This required a fairly high amount of processing and network resources, with 3 multi-core servers and a 20Mbit connectivity completely dedicated to the capture process during the day, from 2pm until 11pm.

Among all the captured information, the previously defined Natural Language Analysis and GeoParsing/GeoReferencing procedures were applied to identify content which was actually relevant to the protest. This step has been performed according to a number of different approaches:

– messages whose geographical origin was located along the areas touched by the protest, at relevant times

– messages explicitly naming places touched by the protest, at relevant times

– messages discussing the protest in one of several possible forms (e.g.: mentioning the protest, its participants, its themes, its path, and more)

This analysis, using a series of different threshold levels to define the level of acceptable quality of the inferred relevancy, which was never placed below 95% for all modalities, allowed to select more than 92000 information elements in the time-frame of the protest.

A series of visualizations were designed to investigate on the results of the process.

A first visualization was designed to show the intensity of communication over time in the various areas of the city. A geo-referenced parametric surface was configured to receive the number of posts in each area of the city as values determining the surface's heights in the matrix of control points. The effect was to create an immediate readability of the locations in which online activity was stronger during the time of the protest.

By superimposing the visualization with the path followed by the protest, it was immediate to understand how the online activity closely followed the protest itself: the march took place both in the physical space and in the digital one.

This form of quantitative, geo-referenced analysis produced evidence of the following two phenomena:

– a high number of people who were physically present at the protest produced digital content and published it on social networks, allowing to observe the impressions, emotions and information as communicated directly from relevant locations at a high level of detail;

– a high number of people who were not physically present at the protest discussed it online, allowing for observation of the general experience of the event.

Then further analysis was performed on the qualitative level, to observe the types of information which could be extracted from the captured streams. This kind of observation was performed using the results of the Natural Language Analysis phase, thus benefiting from the availability of a classification of all information elements according to a classification of emotions and of topics.

The richness of the captured data suggested the possibility to envision, design and implement a series of applicative scenarios.

Given the specific focus on emergent crisis situations in urban contexts, such as those which potentially can take place during protests and revolts, mobile applications and the supporting technological frameworks were designed for the following scenarios/actors:

- a real-time geographical application for public police and security personnel
- a real-time geographical and augmented reality application for protesters
- a real-time geographical application for a fictional type company whose business model is based on the offering of services for these kinds of emergency scenarios

Each application has been thought out according to a dedicated perspective:

- the application for the police forces
    - identification of a series of linguistic templates which would indicate the emergence of specific scenarios which represent dangerous situations or, more in general, situations in which a direct police intervention is required (e.g.: "they're breaking the windshields of the cars" would be among the possible sentences which this part of the system would need to react to and, thus, constructs such as "? breaking ? cars ?" would be a typical part of the linguistic template dictionary used in the platform)
    - identification of rising trends, which might indicate emergent situations which could benefit from the attention of the police forces (e.g.: a sudden rise of messages like "the protest is turning left onto XXX street" would definitely need some attention by police officers, who might decide to intervene in regulating the mutated use of public space);
- the application for the protesters
    - a map and an augmented reality display allow the user to see in real-time what is being communicated in the various directions around the current geographical position
    - several prepared configurations allow the user to see in immediately accessible and understandable ways the spatial distribution of information around own position (e.g.: the colors red and green are used to draw a circle in AR around the user to inform about the presence, in that direction, of messages describing possible situations of danger, such as riot, police charge, injured people; this information would, for example, suggest the user to choose to walk in "green" directions, and to avoid moving towards "red" ones);
    - the user can configure a list of social network users: visual displays constantly show the configured people's positions, thus allowing the user to be constantly aware of their position, thus avoiding getting lost or separated from them, or to establish highly accessible means of spatial communication in emergency scenarios);
- the application for the fictional company
    - a web framework allows the fictional company to setup a curation environment in which to aggregate content harvested in real-time among geo-referenced information published by users on social networks;
    - the framework offers easy tools to observe in real-time the content produced on social networks about a series of strategic themes (paths of protesters in city space, alerting of exceptional events, signals of violence or other dangerous activities;
    - the framework also highlights emerging topics among the real-time expressions of social network users, whose growth in intensity and frequency signals them as interesting-to-observe and, thus, allows to add them among the topics under observation on the city

- map;
- the fictional company's personnel (or software systems) can use these aggregated informations to dynamically create visualizations in which one or more themes are shown and; each grouped representation of this kind (set of layers of manually or automatically curated information) forms a "product" which the company "sells" to various actors, thus realizing its business model;
- to access the offered products/services, users download a smartphone application; when they do, they can choose among the themes aggregated by the fictional company, for example wishing to be alerted of the overall activity relevant to the protest; from that moment that high-quality aggregated information will be shown on a map and around them, using Augmented Reality;
- users can also choose to form a group among other users of the application and including users from supported social networks; in this way they will also see the icons of these users highlighted onto the map and in AR, allowing to know their relative position in real-time and to instantly exchange information.

Simulations of these three platforms produced the following results:

- in a post-event simulation, the police-related application generated around 30000 elements of geo-referenced information classified according to the following themes:
  - violence: around 12000 messages were harvested, of which about 600 allowed to precisely identify the occurrence of violent events;
  - law infringements: around 10000 messages allowed to describe various types of law infringements, such as theft, damage, personal attack, use of weapons of various kinds;
  - abnormal gatherings: in three occasions, unexpected amounts of messages produced in limited geographical areas were identified, allowing to predict the unplanned aggregation of protesters while they were changing the agreed path for the march (evidence of this was confirmed by confronting the stream of news to the timeline of the protest)
  - injuries: around 3000 messages gave evidence of the location of injured people, sometimes including a basic description of the type of injury and an informal description of the state of its gravity
- in a post event simulation, the profiles of 12 random users among the ones who produced information from the protest were selected as users of the protester-oriented application (tests were, thus, performed by using public information and by selecting user connections that were also present at the protest to activate the functionalities of the test; user profiles were not used in the simulation, in which only the visibility of data and user connections was used to form the experience of the simulation; no exceptional event, such as no availability of network connection, has been taken into account in these simulations):
  - some scenarios dealt with the persona of a peaceful protester, wishing to peacefully manifest and march in the city, and desiring to avoid getting hurt or loosing contact with friends and companions; during 3 simulated hours, centered onto the most violent turnouts of the protest, the user benefited from around 2000 real-time information elements allowing to identify the relative position of most violent happenings, and was constantly able to keep in touch with friends that were active on social networks during that timeframe;
  - some other scenarios dealt with the possibility that this application would be used by violent protesters, using it to know the location of police forces inferred by the

- descriptions published on social networks by other users, and to use the violence-related information as a strategic tool to organize groups and interventions; around 12000 real-time information elements were used for this, allowing them to gain substantial strategic edge; companion-related position information allowed this profile to benefit from a practical tool to keep active groups spatially aggregated, and to also immediately visualize the position and direction of movement of other protesters;

- in a post simulation the interface used to curate the real-time information used by the fictional company profile was able to harvest and classify around 10000 high-relevance elements of information, grouped under 6 product/services:

    - violence taking place

    - curiosities (a selection of the most interesting things taking place during the protest, expression of creativity and innovation)

    - dangerous areas

    - joyful events (such as improvised concerts, clown/busker shows and other similar things)

- the spatial messaging features of the fictional company profile have only been lightly tested, verifying their functionality but not yet enough to infer the quality and effectiveness of this practice, requiring a live scenario to fully comprehend its reach.

*VersuS, planet edition*

An enhancement and generalization of the VersuS platform has been recently tested in a prototype which allows to observe several cities at once.

The platform was presented and tested during an italian national radio broadcast using the narrative of a musical journey touching 6 urban contexts (Milan, Berlin, London, Bristol, New York and Philadelphia), with the DJ playing music by artists in the different cities while a web interface allowed listeners to view the real-time information visualizations of those cities. For the event an emotional approach was used, classifying user generated content by emotions organized around the scheme proposed by Robert Plutchik in 1980.

The experiment was closely monitored using a mixture of techniques involving the use of web analytics and direct engagement with listeners through social networks and questions posed during the radio show.

Response has been particularly strong in this occasion. Listeners actively used the platform, constantly inferring meaning and explanations for both the emotional configurations expressed in cites and for the specific messages that, while captured, were being shown on the interfaces.

Listeners autonomously suggested multiple usage scenarios for the platform, also referring to hypothetical scenarios in which these kinds of systems could be used to create participatory governance practices for entire cities. Usage scenarios dedicated to novel entertainment products and services were also often hypothesized, with users declaring their welcoming approach to these kinds of systems being available on their smartphones.

**Conclusions**

The possibility to listen to the ideas, visions, emotions and proposals which are expressed each day by citizens – either explicitly or implicitly by the ways in which they use their cities, workplaces, malls... – suggests the emergence of positive scenarios.

*Harvesting* systems allow us to continuously sense the public discussion and to correlate it to cities, transport systems, infrastructures, architectural spaces, neighborhoods.

"Sensibility Networks" can be established using natural language analysis processes allowing us to "read" cities, for how they are "written" by people, traversing languages and cultures.

Sensor networks can be included in the scenario to record in real-time information about pollution, traffic and the other measurements which shape the ecological, social, administrative and political lives of our cities.

It is possible to create multiple layers of narratives which traverse the city and which allow us to read them in different ways, according to different strategies and tactics, and enabling us to highlight how cities (through their citizens or even on their own, expressing through sensors) express points of view on the environment, culture, economy, transports, energy and politics.

The ubiquitous accessibility of the information about how multiple agencies re-interpret space reveals novel uses for it, thus defining a new *structure* for public space.

The experience of space/time in urban contexts comes out deeply modified, as we progressively mutate our interpretation of presence, space and relation, adding the wide array of usage grammars for space and time to our vocabularies of tools which we use to navigate everything, from maps, to spaces to written text.

Digital information starts contributing to the affordances of the objects, buildings and other things we find in the space around ourselves, as we progressively, pragmatically and *naturally* adopt the idea of having the availability of additional sensorialities which are externalized onto devices and which shape our experience of the world, just as our eyes, ears, fingers...

A mobile phone call can transform a park bench into a temporary, ubiquitous office. A social mapping service can alter our perception of space. An augmented reality system can make visible information on pollution, mobility, energy of the place we are in. A wearable technology can create a new sense connected to remote objects, events, quantities. A real-time digital conversation analysis system can interconnect thoughts, visions, desires and emotions of people and organizations, materializing them onto a novel form of digital space in which identity, privacy and ethics must be redefined.

These methodologies for real-time observation of cities can be described as a form of "ubiquitous anthropology", based on the idea that we can take part in a networked structure shaped as a diffused expert system, capturing disseminated intelligence to coagulate it into a framework for the real-time processing of urban information.

In this context infoaesthetic representations become enablers to enact radical strategies to maximize the accessibility and usability of this information.

Together, all these elements describe something which we might refer to as "ubiquitous user generated search engine", through which citizens become preferential channels for the production of relevant information about themes which are fundamental for our daily lives, giving shape to a scenario in which the concepts of citizenship and political representation can be reinvented, tending towards a vision in which people can be more aware and benefit from added opportunities for action, participating to an environment designed for ubiquitous collaboration and knowledge which is multi-actor, *multi-stakeholder*, in real-time: the city.

**List of Figures**

Figure 1

The Atlas of Rome, the urban screen

Figure 2

VersuS, love VS turin, a moment in the emotions of Turin

Figure 3

City of Rome, sometimes during a research in winter 2011. The image represents a screen capture of a visualization of the messages of citizens of the city of Rome using social networks to discuss the financial crisis in Italy. Red colored dots represent expressions of particular verbal violence.

Figure 4

The city of Rome during the riots of October 15th 2011. The surfaces show the intensity of digital messages exchanged during the riots.

Figure 5

City of Rome, February 15th 2012. Each sphere represents a single social network message sent from 10am to 10:30am from the territory of the city of Rome. Each message is color-coded according to the emotion expressed in the message. The height of the message describes the length of the conversation it generated.

Figure 6

City of Rome, February 15th 2012. Each dot around the circle represents a social network user residing in the city of Rome. 1000 users are shown with their mutual interactions from 2pm to 6pm. The 1000 users are chosen from all users in Rome in order to define a "community": a group of people who intensively exchange information.

**Relevant projects**

ConnectiCity, including the Atlas of Rome, ConnectiCity Neighborhood edition, Architettura rel:attiva.

http://www.artisopensource.net/category/projects/connecticity-projects/

CoS, Consciousness of Streams

http://www.artisopensource.net/category/projects/consciousness-of-streams-projects/

Nuclear Anxiety

http://www.artisopensource.net/category/projects/nuclearanxiety/

Squatting Supermarkets

http://www.artisopensource.net/category/projects/squatting-supermarkets-projects/

The Electronic Man

http://www.artisopensource.net/category/projects/electronicman/

VersuS, the realtime lives of cities

http://www.artisopensource.net/category/projects/versus-projects/

**Links to personal websites, including portfolio, previous work, curriculum**

http://www.artisopensource.net

http://www.fakepress.it